\documentclass{JHEP3}

\usepackage{graphicx}
\usepackage{verbatim}
\usepackage{epsf}
\usepackage{latexsym}
\usepackage{amsmath,amsfonts,amssymb,amsthm}

\newcommand{\bbox}{\lower.2ex\hbox{$\Box$}}

\newcommand{\beq}{\begin{equation}}
\newcommand{\eeq}{\end{equation}}
\newcommand{\bea}{\begin{eqnarray}}
\newcommand{\eea}{\end{eqnarray}}
\newcommand{\ena}{\end{eqnarray}}

\newcommand {\non}{\nonumber}

\newcommand{\Tr}{{\rm Tr}}

\newcommand{\be}{\begin{equation}}
\newcommand{\ee}{\end{equation}}

\preprint{}

\title{\begin{center} 
Metastable Vacua \\
in Superconformal SQCD-like Theories
\end{center}}
\author{
Antonio Amariti$^{1,a}$, Luciano Girardello$^{2,b}$,
Alberto Mariotti$^{3,c,}$,
Massimo Siani$^{2,d}$
\\~
\\
$^1$Department of Physics, University of California\\
San Diego La Jolla, CA 92093-0354, USA
\\~\\
$^2$Dipartimento di Fisica, Universit\`a di Milano Bicocca\\
and \\
INFN, Sezione di Milano-Bicocca,\\ 
piazza della Scienza 3, I-20126 Milano, Italy\\
\\
$^3$
Theoretische Natuurkunde, Vrije Universiteit Brussel \\
and\\
The International Solvay Institutes\\ 
Pleinlaan 2, B-1050 Brussels, Belgium\\
 ~~\\
$^a$\email{antonio.amariti@physics.ucsd.edu} \\
$^b$\email{luciano.girardello@mib.infn.it} \\
$^c$\email{alberto.mariotti@vub.ac.be} \\
$^d$\email{massimo.siani@mib.infn.it}
}

\abstract{ 
We study dynamical supersymmetry breaking in vector-like superconformal 
$\mathcal{N}=1$ gauge theories. We find appropriate deformations of the superpotential to overcome the problem of the instability of the non supersymmetric vacuum. 
The request for long lifetime translates into constraints on the physical couplings which in this regime can be controlled through efficient RG analysis.}
\begin{document}

\section{Introduction}
In the last few 
years many models of 
metastable dynamical supersymmetry breaking (DSB) 
based on the ISS breakthrough \cite{Intriligator:2006dd}
have been proposed (see \cite{Intriligator:2007cp} and references therein).
Usually in DSB the strong dynamics 
jeopardizes the calculability of 
the model.
The novelty of the approach of ISS relies in
describing the low energy theory by the Seiberg dual phase
\cite{Seiberg:1994pq,Intriligator:1995au}
which is weakly coupled in the IR.
For a $\mathcal{N}=1$ $SU(N_c)$ supersymmetric gauge theory
with $N_f> N_c+1$ flavors 
the
low energy physics can be equivalently described
by a different \emph{magnetic} $SU(N_f-N_c)$ gauge group with $N_f$ flavors
and a singlet.
Furthermore if $N_f<2 N_c$,
the $SU(N_c)$ gauge group is strongly coupled whereas 
the dual magnetic gauge group $SU(N_f-N_c)$ is weakly coupled
in the IR.

The ISS model is based on  SQCD 
with $N_c+1<N_f<3/2 N_c$ and small masses 
for the quarks. In this window the 
dual gauge theory at low energy 
flows to an IR free fixed point.
This theory breaks supersymmetry at tree level
in the small field region.
In this region the strong dynamics 
effect are safely negligible and perturbation
theory is reliable.
The supersymmetric vacua 
are recovered in another region 
of the field space, namely at large vevs.
The analysis shows that the supersymmetry breaking
vacuum is metastable, and the lifetime 
of this state can be made parametrically large by
tuning the scales of the theory.

In principle the same mechanism is applicable
in the conformal window 
if $3/2 N_f < N_c < 2 N_f$, where there
is a weakly interacting fixed point. In 
\cite{Intriligator:2006dd} the authors
showed that in such window the non supersymmetric vacuum
is unstable to decay
because the strong dynamics effects are relevant
and not negligible 
around the origin of the field space.
Indeed the bounce action between the non supersymmetric vacuum and
the supersymmetric one is not parametrically large, and the lifetime is short.
Recent studies for realizing metastable vacua
in the conformal window has been done in \cite{Izawa:2009nz}.

In this paper we investigate this problem more deeply, 
and we find a
viable model of metastable supersymmetry breaking in the 
conformal regime of a SQCD like theory.

We start our analysis by revisiting the ISS model in the conformal window,
studying the RG evolution of the couplings and of the bounce
action.
The lifetime of the non supersymmetric vacuum 
is proportional to the ratio between the IR supersymmetry breaking
scale and the IR holomorphic scale.
We find that
this ratio depends only on the 
gauge coupling calculated at the conformal fixed point.
This shows that the lifetime of the vacuum cannot be 
parametrically large below the IR scale at which the theory 
exits from the conformal regime.

Nevertheless, we argue that by adding some deformations
the metastable vacua can still exist in the conformal window. 
We propose a deformation of the ISS model, by
adding a small number of massive quarks
and some new singlets in the dual description
of massive SQCD.
This model is  a $SU(N)$ SCFT
dual to the SSQCD 
defined in \cite{Barnes:2004jj}
with some relevant deformations.
When these deformations are small, the theory
is approximately a CFT.
In this approximate CFT regime this theory is interacting, 
and we restrict to the weakly coupled 
window such that the perturbative analysis is reliable.
This model can evade the argument of ISS because the new 
massive fields modify the non perturbative superpotential and thus
the supersymmetric vacuum.
As a consequence the
bounce action has a parametrical behavior in terms of
the relevant deformations.  
The lifetime can be large if we impose some constraints
on the physical couplings at the CFT exit scale.

Differently from the IR free case, in which the low energy theory
is free, in this case the model is interacting. The anomalous 
dimensions of the fields are not zero, and the Kahler potential is renormalized. 
This implies that  the physical couplings undergo RG evolution 
in the approximate CFT regime.
The constraints for the stability of the non supersymmetric vacuum
have to be imposed on the physical IR couplings after RG evolution.
These translate in conditions for the UV couplings and for
the duration of the approximate CFT regime.
We then look for the allowed region of UV couplings such that the 
bounds on the lifetime of the vacuum,
imposed in the IR, are satisfied.

We argue that metastable vacua are common in the conformal window,
and we give a procedure to find other models. The basic requirement is that there
must be a regime of parameters and ranks such that the supersymmetric
vacua are far away in the field space, and that the bounce action 
is a function of the relevant deformations.
As in  SSQCD, which is the simplest example,
a richer set of relevant deformations than in
massive SQCD is necessary. 

The paper is organized as follows. In Section \ref{Pincopalla} we 
discuss the obstructions to the existence of 
metastable vacua in SQCD in the conformal window,
and we introduce the analysis of the RG evolution for the couplings
and the holomorphic scale. 
In the main Section \ref{mainsec} 
we outline our strategy for the search of metastable vacua 
by studying the SSQCD model appropriately deformed. 
The key point just relies on the features of super CFT, 
where RG analysis and determination of anomalous 
dimensions are feasible.
In Section \ref{dafare} 
we discuss the generalization of our analysis to $\mathcal{N}=1$
SCFTs. In Section \ref{disco} we conclude. In the Appendix \ref{appbounce}
we study the RG flow associated to the bounce action.
In the Appendix \ref{appanto} 
we review the Seiberg duality in SSQCD and discuss
the origin of the relevant 
couplings.
\\
\\
While we were completing this paper, the work 
\cite{Yanagida:2010aa} appeared 
which has some overlap with our results.

\section{The case of SQCD}
\label{Pincopalla}
In the original paper \cite{Intriligator:2006dd} the authors studied
a $SU(N_c)$ gauge theory with $N_f$ flavors of quarks
charged under an $SU(N_f)^2$ flavor symmetry
broken to $SU(N_f)$ by the superpotential 
\be 
W = m Q \tilde Q
\ee
where the mass $m$ is much smaller than the holomorphic scale of the
theory $\Lambda$.
In the window $N_c+1 < N_f$ this theory
admits a
dual description in term of a \emph{magnetic} gauge group 
$SU(\tilde N)=SU(N_f-N_c)$,
$N_f$ magnetic quarks $q$ and $\tilde q$ and the \emph{electric} meson
$N=Q \tilde Q$ normalized to have mass dimension one.
The dual superpotential reads
\be \label{WDUALE}
W_m = - h \mu^2 N+ h N q \tilde q + \tilde N
\left(        \tilde \Lambda^{\tilde b} h^{N_f} \det N  \right)^{\frac{1}{\tilde N}}
\ee
where we introduced the marginal coupling $h$ and the holomorphic scale
of the dual theory $\tilde \Lambda$, and we added the non perturbative
contribution due to gaugino condensation.
From now on we set $h=1$.
The holomorphic scales $\Lambda$ and $\tilde \Lambda$ 
are related by a scale matching relation \cite{Intriligator:1995au}.
The one loop beta function coefficient is $\tilde b=3 \tilde N -N_f=2 N_f-3 N_c$.

In the range $N_c+1 < N_f < 3/2 N_c$, this
theory has a supersymmetry breaking vacuum at $N=0$, 
with non zero vev for the quarks.
The supersymmetric vacuum is recovered in the large field region for $N$.
The parametrically long distance between the two vacua guarantees
the long life time of the non supersymmetric one.

The metastable non supersymmetric 
vacua found in the magnetic free window of massive SQCD
are destabilized in the conformal window $3/2 N_C < N_f < 3N_C$.  This
fact is based on the observation that the non perturbative
superpotential in (\ref{WDUALE}) is not negligible in the small field
region, as instead it happens in the magnetic free window.

Here we study more deeply this problem.  
In general, in the presence of relevant deformations the
conformal regime is only approximated. If these deformation are small
enough there is a large regime of scales in which the theory flows to
lower energies remaining at the conformal fixed point.  The physical
couplings vary along the RG flow because of the wave function 
renormalization of the fields, until the theory exits  from the conformal
regime.  Below this scale the theory is IR free and the renormalization
effects are negligible. 

We study the RG properties of the ISS model in the conformal
window by using a canonical basis for the fields.
Flowing from a UV scale $E_{UV}$ to an IR scale $E_{IR}$
the fields are not canonically normalized anymore, and we
have to renormalize them by the wave function renormalization 
$Z_i(E_{IR},E_{UV})$, namely
$\phi_i^{IR}=\sqrt Z_i \phi_i^{UV} $.
In terms of the renormalized fields the Kahler potential is 
canonical. 
The
couplings appearing in the superpotential 
undergo RG evolution, and are
the physical couplings. 
In this way the coupling $\mu_{IR}$
of the IR superpotential becomes
\be
\mu_{IR} = \mu_{UV} Z_N(E_{IR},E_{UV})^{-\frac{1}{4}}
\ee
The holomorphic scale that appears in the superpotential 
is unphysical in the conformal window and it is defined as
\be
\label{holoscala}
\tilde \Lambda= E e^{-\frac{8 \pi^2}{g_*^2 \tilde b}}
\ee
where E is the RG running scale, and $g_*$ is the gauge coupling
at the superconformal fixed point.
In the canonical basis $\tilde \Lambda$ is rescaled 
as well
during the RG conformal evolution
as \cite{ArkaniHamed:1997mj,ArkaniHamed:1997ut,
Poland:2009yb}
\be
\tilde \Lambda_{IR} = \tilde  \Lambda_{UV} \frac{E_{IR}}{E_{UV}} 
\ee
In the ISS model the two possible sources of breaking
of the conformal invariance
are the masses of the fields at the non supersymmetric vacuum
and the masses of the fields at the supersymmetric vacuum.
We define the
CFT exit scale as $E_{IR}=\Lambda_c$. In this model
this scale is necessarily set by the masses of the fields
at the supersymmetric vacuum, which are proportional to the vev
of the field  $N$. In fact  by setting
\be \label{vevmes}
\Lambda_c \equiv 
\langle N \rangle_{susy} = \mu_{IR}\left(\frac{\mu_{IR}}{\tilde \Lambda_{IR}}\right)^{\frac{\tilde b}{N_f-\tilde N}}
\ee
the physical mass at this scale 
results
\be
\mu_{IR} = \Lambda_c e^{- \frac{4 \pi^2}{{g*}^2 \tilde N}} \ll \Lambda_c
\ee
Hence the assumption that $\langle N \rangle_{susy}$ stops the conformal
regime is consistent.
The opposite case, with 
$\Lambda_c \equiv \mu_{IR} \gg \langle N \rangle_{susy}$
cannot be consistently realized.

The bounce action at the scale $\Lambda_c$ is
\be \label{bbb}
S_B \sim \left( 
\frac{\mu_{IR}}{\tilde \Lambda_{IR}} 
\right)^{\frac{4 \tilde b}{N_f-\tilde N}}
\sim e^{ \frac{16 \pi^2}{{g}_{*}^2 \tilde N}}
\ee
This bounce is not parametrically large
and it depends only
on the coupling constant $g_*$ at the fixed point.
In general, as we shall see in the appendix \ref{appbounce}, the 
bounce action is not RG invariant, but it runs during the 
RG flow.
In this case $S_B$ at the CFT exit scale 
only depends on the ratio of the two relevant scales 
in the theory which is the RG invariant coupling constant.

In general, by adding other deformations, the bounce 
action is not RG invariant anymore and we have to take care
about its flow. 
In some cases, the lifetime of a vacuum decreases
as we flow towards the infrared.
In the next section, by adding new massive quarks to the 
ISS model, we show that long living metastable vacua exist in
the conformal window.

\section{Metastable vacua by adding relevant deformations}
\label{mainsec}
In this section we describe our proposal for realizing metastable 
supersymmetry breaking 
in the conformal window of $N=1$ SQCD-like theories. 
The key point is the addition of massive quarks in the dual 
magnetic description. This introduces a new mass scale
that controls the distance in the field space of the supersymmetric vacuum.

We consider the magnetic description of the ISS model 
of 
the previous section. 
We add a new set of massive fields $p$ and $\tilde p$ charged under a
new $SU(N_f^{(2)})$ flavor symmetry. 
We also add new bifundamental fields 
$K$ and $L$
charged under 
$SU(N_f^{(1)} ) \times SU(N_f^{(2)})$. 
The added number of flavors is such that
$3/2 \tilde N <N_f^{(1)} + N_f^{(2)} < 3 \tilde N$.
The superpotential of the model is
\be \label{spotimpo}
W = K p \tilde q + L \tilde p q + N q \tilde q + \, \rho p \, \tilde p -\mu^2 N
\ee
and the field content is summarized in the Table \ref{tabella1}.
\begin{table}[htbp]
\begin{center}
\begin{tabular}{c||c|c|c}
&$N_f^{(1)}$&$N_f^{(2)}$&$\tilde N$ \\
\hline 
$N$&$N_f^{(1)} \otimes N_f^{(1)} $&1&1\\
$q + \tilde q$&$\bar{N}_f^{(1)}\oplus N_f^{(1)}$&1&$\tilde N\oplus \bar{\tilde N}$\\
$p+\tilde p$&1&$\bar{N_f}^{(2)}\oplus N_f^{(2)}$&$\tilde N \oplus \bar{\tilde N}$\\
$K + L $&$\bar{N_f}^{(1)}\oplus N_f^{(1)}$&$N_f^{(2)} \oplus \bar{N_f}^{(2)}$&1\\
\end{tabular}
\caption{Matter content of the dual SSQCD}
\label{tabella1}
\end{center}
\end{table}
This model is the dual description of the SSQCD 
studied in \cite{Barnes:2004jj}, deformed by two relevant operators.
In the appendix \ref{appanto} we show the Seiberg dual electric description of this
theory, and we discuss a mechanism to dynamically generate
the mass term for the new quarks.

In the rest of this section we  
show that in the case of $N_f^{(1)}>\tilde N$
there are ISS like metastable supersymmetry 
breaking vacua if we are near the IR free 
border of the conformal window, 
i.e. $N_f^{(1)}+ N_f^{(2)} \sim 3 \tilde N$.

We  shall
work in the window between the number of flavor and 
the number of colors
\be
\label{window}
2 \tilde N <N_f^{(1)} + N_f^{(2)} < 3 \tilde N
\ee
such that the gauge group is weakly coupled and we can rely on
the perturbative analysis.

\subsection*{The non supersymmetric vacuum}
The non supersymmetric vacuum 
is located near the origin of the field space where
the superpotential (\ref{spotimpo}) 
can be studied perturbatively.
Neglecting the non perturbative dynamics requires
some bounds on the parameters
$\rho$ and $\mu$.
In the rest of the paper we will see that these bounds can be consistent 
with the  running of the coupling constants.

Tree level supersymmetry breaking is possible if 
\be
N_f^{(1)}> \tilde N  \qquad \Rightarrow \qquad 2 \tilde N > N_f^{(2)}
\ee
where the second inequality follows from (\ref{window}).
The equation of motion for the field $N$ breaks supersymmetry
through the rank condition mechanism. 
We solve the other equations of motion and we find the 
non supersymmetric vacuum
\begin{eqnarray}
&&
q=\left(\begin{array}{c}
\mu+\sigma_1 \\
\phi_1\end{array}
\right) \quad
\tilde q = ( \begin{array}{cc} \mu+\sigma_2&\phi_2 \end{array})\quad
N=\left( \begin{array}{cc}
\sigma_3&\phi_3\\
\phi_4 &X
\end{array}\right)\nonumber \\
&&
p=\phi_5\quad \tilde p = \phi_6\quad 
L=(\begin{array}{cc}\phi_7&\tilde Y\end{array})
\quad K=\left(
\begin{array}{c}
\phi_8\\ Y \end{array} \right)
\end{eqnarray}
where we have also inserted the fluctuations around the minimum,
$\sigma_i$ and $\phi_i$. The fields
$X$, $Y$ and $\tilde Y$ are pseudomoduli.
The infrared superpotential is
\begin{eqnarray}
W_{\text{IR}}&=&  X \phi_1 \phi_2 -\mu^2 X +\mu(\phi_1 \phi_4 + \phi_2 \phi_3) +
\mu(\phi_5 \phi_8 + \phi_6 \phi_7) \non \\&+& Y \phi_2 \phi_5 + \tilde Y \phi_1 \phi_6
+\rho \phi_5 \phi_6
\end{eqnarray}
In the limit of small $\rho$, this is the same superpotential studied in \cite{Franco:2006es}. 
This superpotential corresponds to the one studied in
\cite{Amariti:2009tu} in the $R$ symmetric limit.
The fields $X$, $Y$ and $\tilde Y$ are stabilized by one loop corrections
at the origin with positive squared masses.

\subsection*{The supersymmetric vacuum}
We derive here the low energy effective action 
for the field $N$, and we recover
the supersymmetric vacuum in the large field region.
The supersymmetric vacuum is characterized 
by a large expectation 
value for $N$. 
This vev gives mass to the quarks $q$ and $\tilde q$ and we
can integrate them out at zero vev. Also the quarks $p$ and $\tilde p$
are massive and are integrated out at low energy.
The scale of the low energy theory $\Lambda_L$ is related to the holomorphic scale
$\tilde \Lambda$ via the scale matching relation
\begin{equation}
\Lambda_L^{3\tilde N}
=
\tilde \Lambda^{3\tilde N-N_f^{(1)}-N_f^{(2)}} \det \rho \det N
\end{equation} 
The resulting low energy theory is 
$\mathcal{N}=1$ SYM plus a singlet, with effective superpotential
\begin{equation}
W = - \mu^2 N + \tilde N
(\tilde \Lambda^{3\tilde N-N_f^{(1)}-N_f^{(2)}} \det \rho\det N)^{1/\tilde N}
\end{equation}
where the last term is the gaugino condensate.
By solving the equation of motion for $N$
we find the supersymmetric vacuum 
\be
\langle N \rangle_{susy}= \frac
{\mu^{
\frac{2 \tilde N }{N_f^{(1)}-\tilde N}}}{
\tilde \Lambda^{\frac{3\tilde N -N_f^{(1)} -N_f^{(2)}}{N_f^{(1)}-\tilde N}}
\rho^{\frac{N_f^{(2)}}{N_f^{(1)}-\tilde N}} }
\ee

\subsection*{Lifetime}
The lifetime of the 
non supersymmetric vacuum is controlled
by the bounce action to the supersymmetric vacuum.
In this case, the triangular approximation \cite{Duncan:1992ai}
is valid and the action can be approximated as
$S_B \simeq(\Delta \Phi)^4/(\Delta V)$.
If we estimate $\Delta \Phi \sim \langle N \rangle_{susy}$ and $\Delta V \sim \mu^4$
we obtain
\begin{equation} \label{sbb}
S_B = \left(\frac{\tilde \Lambda}{\rho}\right)^{\frac{4 N_f^{(2)}}{N_f^{(1)}-\tilde N }       } 
\left(\frac{\mu}{\tilde \Lambda}\right)^{\frac{ 12 \tilde N - 4N_f^{(1)}}{N_f^{(1)}-\tilde N}  } 
\end{equation}
This expression is not automatically very large since $\mu \ll \tilde \Lambda$.
However, we can impose the following bound on $\rho$
\begin{equation}
\label{primob}
\langle N \rangle_{\text{susy}} \gg \mu \quad \rightarrow  \quad  \rho \ll \tilde \Lambda 
\left( \frac{\mu}{\tilde \Lambda} \right)^{(3\tilde N-N_f^{(1)})/N_f^{(2)}}
\end{equation}
If this bound is satisfied,
the supersymmetric and the non supersymmetric vacua are far away apart 
in the field space
and
the non perturbative terms can be 
neglected at the supersymmetry breaking scale.
This differs from the ISS model in the conformal window.
In that case
the non-perturbative effects became important at the 
supersymmetry breaking scale.
The bounce action
was proportional to the 
gauge coupling constant at the fixed point and it was 
impossible to make it parametrically long. 
The introduction of the new mass scale $\rho$ allows a solution
to this problem.

The bound (\ref{primob}) should be imposed on the 
IR couplings at the CFT exit scale $E_{IR} =\Lambda_c$. 
In this case we have a new possible source of CFT breaking, namely
the relevant deformation $\rho$. 
However we look for a regime of couplings such that the 
CFT exit scale is set by the supersymmetric vacuum scale,
i.e. 
$\Lambda_c = \langle  N \rangle_{\text{susy}} \gg \mu_{IR},\rho_{IR}$.
The scale $\Lambda_c$ is 
\be \label{telotronco}
\Lambda_c =\langle  N \rangle_{\text{susy}}= \tilde \Lambda_{IR} \left(\frac{\mu_{IR}}{\tilde
\Lambda_{IR} }\right)^{\frac{2\tilde N}
{N_f^{(1)}-\tilde N}} 
\left( \frac{\tilde \Lambda_{IR} }{\rho_{IR}}\right)^{\frac{N_f^{(2)}}{N_f^{(1)}-\tilde N}} 
\ee
At this scale we define $\epsilon_{IR}$ as the ratio between the 
IR masses $\rho_{IR}$ and $\mu_{IR}$ and we demand that
\be
\label{epsilon}
\epsilon_{IR}=\frac{\rho_{IR}}{\mu_{IR}} \ll 1
\ee
Rearranging (\ref{telotronco}) for $\mu_{IR}$ and $\rho_{IR}$ we have
\bea \label{provaci!!}
\mu_{IR}= \Lambda_c e^{-\frac{8 \pi^2}{g_{*}^{2} (2 \tilde N - N_f^{(2)})}} 
\epsilon_{IR}^{\frac{N_f^{(2)}}{2\tilde N - N_f^{(2)}}} \ll \Lambda_c\non \\
\rho_{IR}= \Lambda_c e^{-\frac{8 \pi^2}{g_{*}^{2}(2 \tilde N - N_f^{(2)})}} 
\epsilon_{IR}^{\frac{2 \tilde N}{2\tilde N - N_f^{(2)}}} \ll \Lambda_c \non \\
\eea
This shows that requiring $\epsilon_{IR} \ll 1$ is consistent with the
CFT exit scale to be $\langle N \rangle_{susy}$.

By substituting (\ref{telotronco}) and (\ref{provaci!!}) in 
(\ref{sbb}), the bounce action becomes
\be
S_B = 
\frac{e^{ \frac{32 \pi^2}{g_{*}^2 (2\tilde N-N_f^{(2)})}}}
{\epsilon_{IR}^{ \frac{4 N_f^{(2)}}{2 \tilde N-N_f^{(2)}}}}
\ee
and in the limit $N_f^{(2)} \to 0$ it reduces to the one computed in the (\ref{bbb}).
Here the bounce is not only proportional to a numerical factor
depending on $g_{*}^{2}$, but there is also a parameter, relating the ratios of
the physical masses $\rho_{IR}$ and $\mu_{IR}$ at the CFT exit scale.
The bounce action can be large if $\epsilon_{IR} \ll 1$, providing a parametrically large 
lifetime for the non supersymmetric vacuum.

Using the RG evolution equations the bound $\epsilon_{IR} \ll 1$
translates in constraints
on the UV masses $\rho_{UV}$ and $\mu_{UV}$ at the UV scale.
These masses are relevant perturbations
and their ratio must be small along the RG flow.

\subsection*{RG flow in the approximate conformal regime}

The relevant coupling constants run from  
$E_{UV}$ to $E_{IR}=\Lambda_c$. 
We require that these terms are so small in the UV
to be considered
as perturbations of the CFT, i.e.
$\rho_{UV}, \mu_{UV} \ll \Lambda_{UV}$.

The ratio $\epsilon_{UV}$ given at the scale $E_{UV}$ 
runs as the coupling constants down to $\Lambda_c$.
We now study the evolution of this ratio.
The requirement
 of long lifetime of the metastable vacuum  (\ref{primob})
corresponds to 
 $\epsilon_{IR}\ll 1$ and it constrains both
$\epsilon_{UV}$ 
and the duration of the approximate conformal regime, $\Lambda_{c}/E_{UV}$.

The running of the relevant couplings in the conformal windows is parameterized by
the equations
\bea
&&\rho_{IR} = \rho_{UV} 
Z_p( \Lambda_c,E_{UV})^{-1/2}
Z_{\tilde p}( \Lambda_c,E_{UV})^{-1/2}
\\
&&\mu_{IR} = \mu_{UV} 
Z_N( \Lambda_c,E_{UV})^{-1/4}
\eea
The wave function renormalization 
$Z$ is obtained by integrating  the equation
\be \label{defgamma}
\frac{\text{d} \log Z_i}{\text{d} \log E} = -\gamma_i
\ee
from $E_{UV}$ to $\Lambda_c$, 
where $\gamma_i$ is constant in the conformal regime,
and it reads
\be
Z_\phi ( \Lambda_c,E_{UV})=
\left(
\frac{\Lambda_c}{E_{UV}}
\right)^{-\gamma_{\phi_i}}
\ee
The physical couplings at the CFT exit scale are
\be
\rho_{IR} = \rho_{UV}
\left(
\frac{\Lambda_c}{E_{UV}}
\right)^{\gamma_{p}},
\quad \quad \quad
\mu_{IR} = \mu_{UV}
\left(
\frac{\Lambda_c}{E_{UV}}
\right)^{\gamma_{N}/4}
\ee
where we have used the relation $\gamma_p=\gamma_{\tilde p}$.
Along the flow from $E_{UV}$ to $\Lambda_c$ the coupling
$\mu_{IR}$ is suppressed, because
$\gamma_N>0$, while $\rho_{IR}$ becomes larger, because $\gamma_p<0$.

The ratio $\epsilon$ evolves as
\be
\epsilon_{IR} =\epsilon_{UV} \left(\frac{\Lambda_c}{E_{UV}} \right)^{\gamma_p-\gamma_N/4}
\ee
and we demand that it is $\epsilon_{IR} \ll 1$
in order to satisfy the stability constraint for the non 
supersymmetric vacuum.
The flow from $\epsilon_{UV}$ to $\epsilon_{IR}$ depends
on $\Lambda_c /E_{UV}$ and on the anomalous dimensions. 
The precise relation between $\epsilon_{UV}$ and $\epsilon_{IR}$
is found by calculating the exact value of $\gamma_p$ and $\gamma_N$.
The anomalous dimensions of the fields $\phi_i$ 
are obtained from the relation $\Delta_i = 1 + \gamma_i/2$
where $\Delta_i = \frac{3}{2} R_i$. The $R$ charges can be computed
by using a-maximization.

The a-maximization procedure, defined in \cite{Intriligator:2003jj},
shows that in SCFT 
the correct $R$-charge
at the fixed point is found by maximizing the function
\begin{equation}\label{atrial}
a_{trial} (R) = \frac{3}{32} \left(3 \Tr{R^3} - \Tr R\right)
\end{equation}
The $R$-charges in (\ref{atrial}) are all the non anomalous 
combinations of the $R_0$
charges under which the supersymmetry generators have charge $-1$ and
all the other flavor symmetries commuting with the supersymmetry generators.  
The $\Tr(R^3)$ and $\Tr(R)$ are the coefficients of the gauge anomaly 
and gravitational anomaly.
The $R$-charges that maximize (\ref{atrial}) are the $R$ charges appearing 
in the superconformal algebra.

The $R$ charge assignment has to satisfy 
the anomaly free condition and the constraint that the superpotential
couplings should be marginal.
These conditions are
\be
\tilde N + N_f^{(1)} (R[q]-1) + N_f^{(2)} (R[p]-1)=0, \quad  \,R[p] + R[q] + R[L] =2, \quad
R[N] + 2 R[q]=2 
\ee
where the symmetry enforces $R[q]=R[\tilde q]$, $R[p]= R[\tilde p] $ and $R[K]=R[L]$.
The $a_{trial}$ function that has to be maximized is
\bea
a_{trial} &=& \frac{3}{32} \left( 2 N_f^{(1)} \tilde N \left( 3(R[q]-1)^3 - R[q] +1 \right)
+ 2 N_f^{(2)} \tilde N \left( 3(R[p]-1)^3 - R[q] +1 \right) \right. 
 \non \\
&+& \left. 2 N_f^{(1)}  N_f^{(2)} \left( 3(R[L]-1)^3 - R[L] +1 \right)
+ N_{f}^{(1) \, 2} \left( 3(R[N]-1)^3 - R[N] +1 \right) + 2 \tilde N^2
\right) \non \\
\eea
By defining $ R[N]= 2 y$ we have $R[q] = 1-y$. The other $R$ charges are
\be
R[p] = \frac{1}{n} (n-x+y), \quad
R[L] = y+\frac{x-y}{n} 
\ee
where $n=\frac{N_f^{(2)}}{N_f^{(1)}}$ and  $x=\frac{\tilde N}{N_f^{(1)}}$.
We can simplify the $a$ maximization in terms of the only variable y,
obtaining \small
\bea \label{cicciogay}
&&y_{\text{max}}=
\frac{-3 \! \left(\!n+n^3\!\right)\!+\!3 (\!-\!1\!+\!n\!)^2\! x\!-\!3\! x^2\!\!+\!
\!\sqrt{\!n^2\! \left(\!n^4\!-\!8 n (\! x \! -\! 1 \!)\!+\!8 n^3 (\!x\!-\!1\!)\!+\!9 (\!x\!-\!1\!)^4\!-\!\!6 n^2 (1\!+\!3 (x\!-\!2) x)\right)}}{3 \left(1-n \left(3+n+n^2\right)+\left(-1+n^2\right) x\right)}\non \\&&
\eea
\normalsize

\begin{figure}[!b]
 \begin{minipage}[b]{6cm}
   \centering
   \includegraphics[width=5cm]{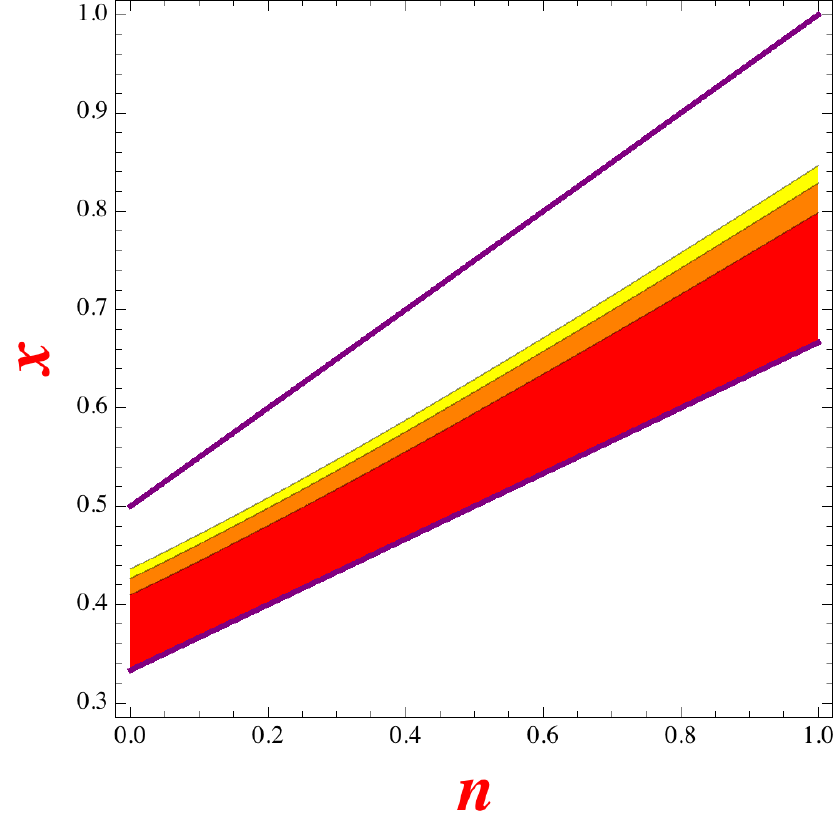}
   \caption{\tiny {$\frac{
\rho_{UV}}{\mu_{UV}}=\!10^{-2},
\frac{\Lambda_{c}}{E_{UV}}=10^{-4}$
\label{primafig}
}}
 \end{minipage}
 \ \hspace{2mm} \hspace{3mm} \
 \begin{minipage}[b]{6cm}
  \centering
   \includegraphics[width=5cm]{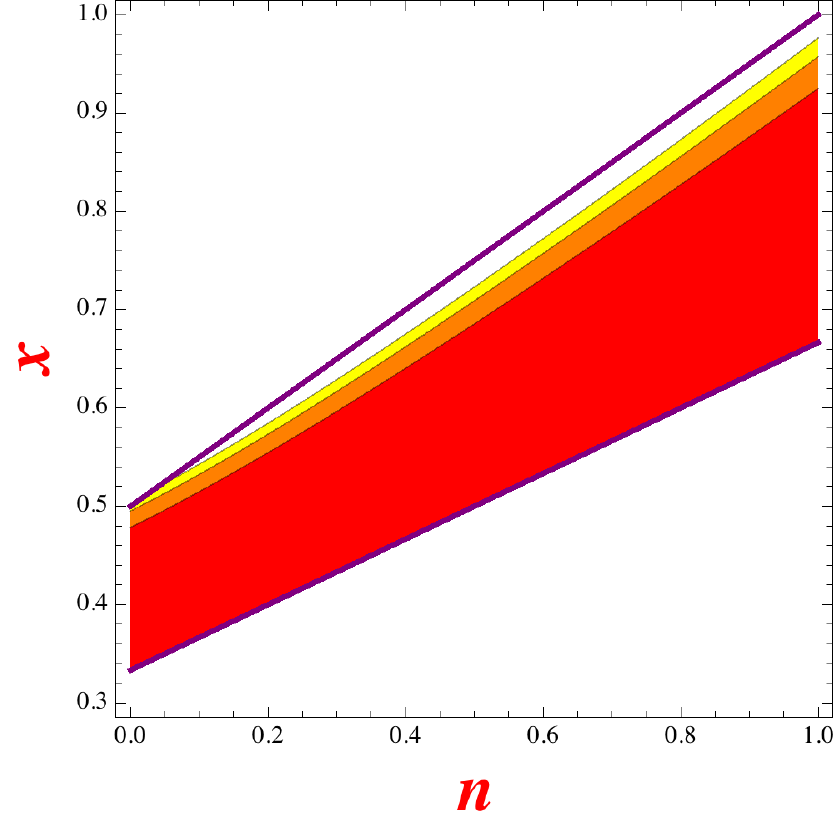}
 \caption{\tiny {$\frac{
\rho_{UV}}{\mu_{UV}}=\!10^{-4},
\frac{\Lambda_{c}}{E_{UV}}=10^{-4}$
}}
 \end{minipage}
  \\~\\~\\
   \begin{minipage}[b]{6cm}
   \centering
   \includegraphics[width=5cm]{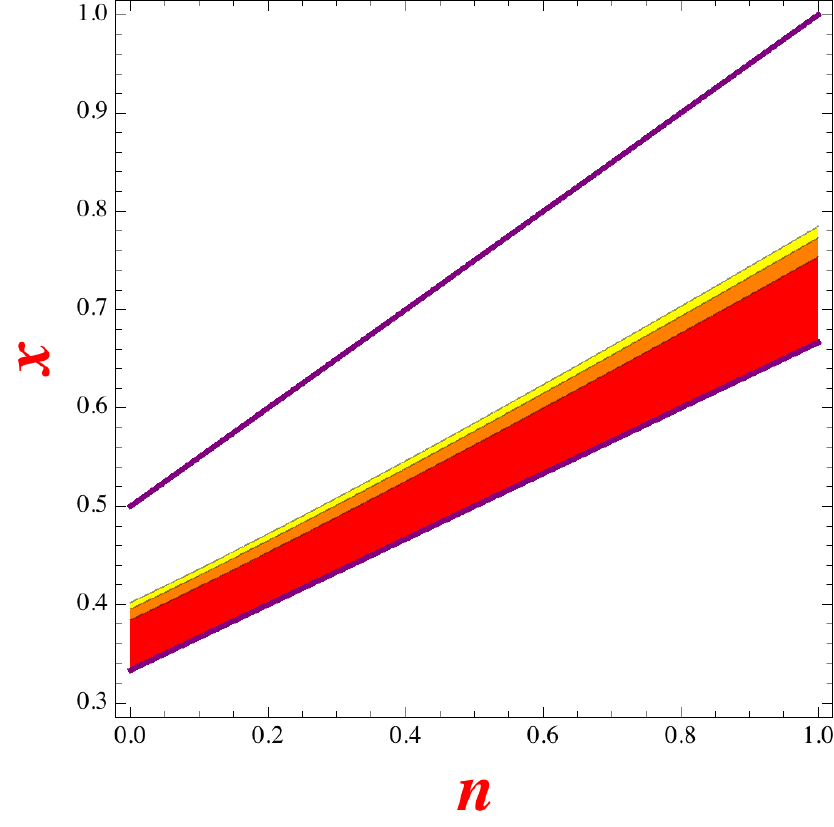}
\caption{\tiny {$\frac{
\rho_{UV}}{\mu_{UV}}=\!10^{-2},
\frac{\Lambda_{c}}{E_{UV}}=10^{-6}$
}}
 \end{minipage}
 \ \hspace{2mm} \hspace{3mm} \
 \begin{minipage}[b]{6cm}
  \centering
   \includegraphics[width=5cm]{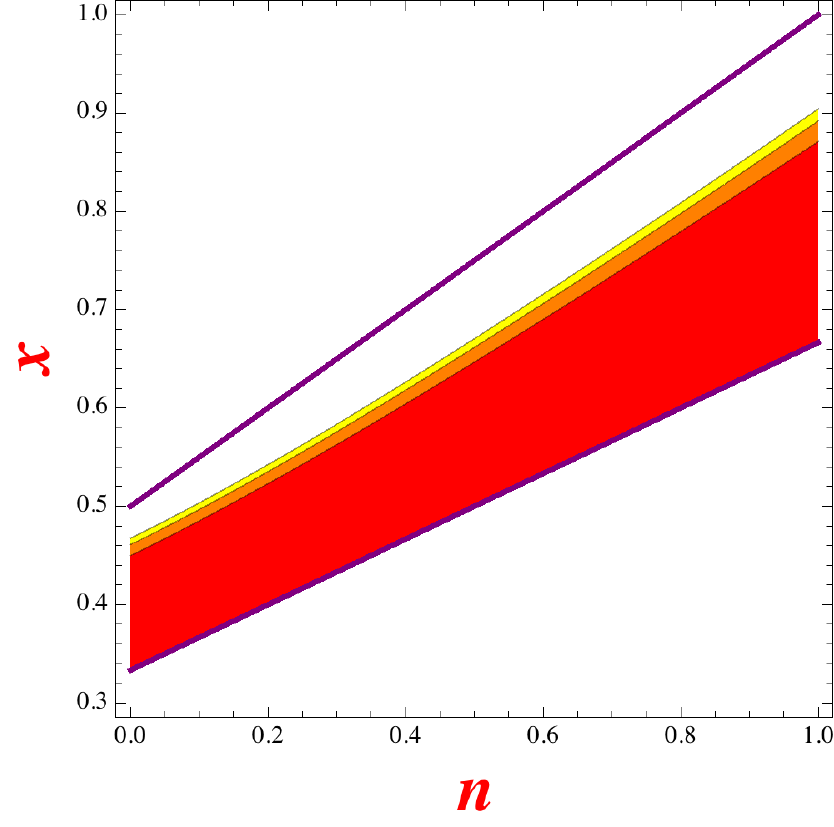}
\caption{\tiny{$\frac{
\rho_{UV}}{\mu_{UV}}=\!10^{-4},
\frac{\Lambda_{c}}{E_{UV}}=10^{-6}$
}}
 \end{minipage}
 \\~\\~\\
 \begin{minipage}[b]{6cm}
   \centering
   \includegraphics[width=5cm]{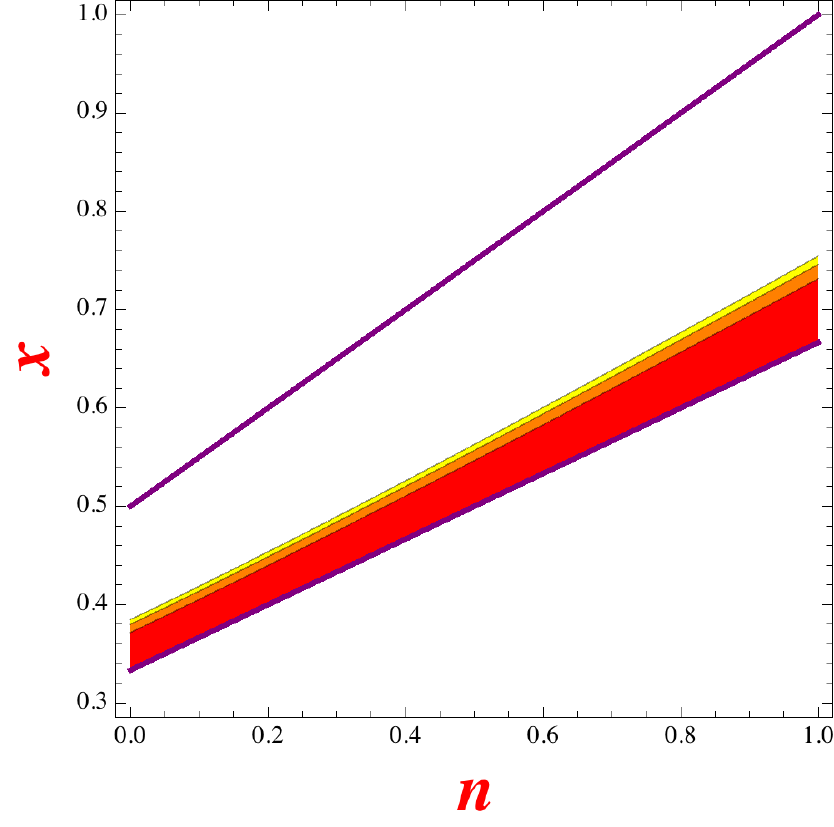}
\caption{\tiny {$\frac{
\rho_{UV}}{\mu_{UV}}=\!10^{-2},
\frac{\Lambda_{c}}{E_{UV}}=10^{-8}$
}}
 \end{minipage}
 \ \hspace{2mm} \hspace{3mm} \
 \begin{minipage}[b]{6cm}
  \centering
   \includegraphics[width=5cm]{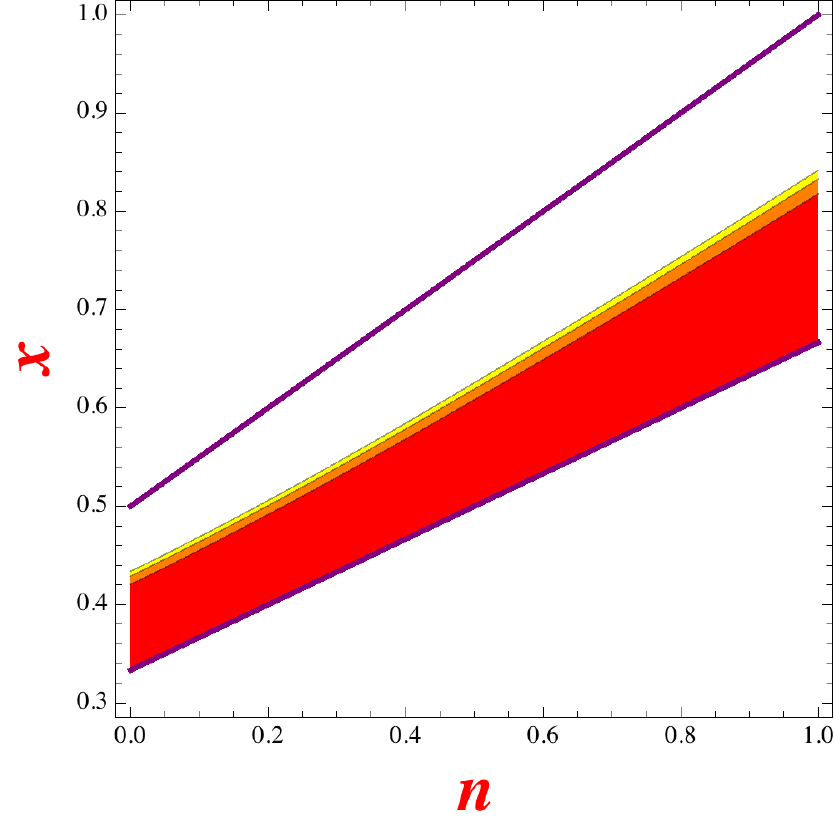}
\caption{\tiny {$\frac{
\rho_{UV}}{\mu_{UV}}=\!10^{-4},
\frac{\Lambda_{c}}{E_{UV}}=10^{-8}$
}}
\label{ultimafig}
 \end{minipage}
  \end{figure}

  Once we know the anomalous dimensions 
and once we fix the duration of the 
approximate conformal regime
we can see
 what is the bound to impose
on the UV ratio $\epsilon_{UV}=\rho_{UV}/\mu_{UV}$
such that 
\be
\epsilon_{IR}=\epsilon_{UV} 
\left(\frac{\Lambda_c}{E_{UV}} \right)^{\frac{3}{2n} (n-2 x+ 2 y - y n)} \ll 1
\ee
In the Figures \ref{primafig}-\ref{ultimafig} we have plotted some region of the ranks 
$x$ and $n$  by fixing $\epsilon_{UV}$
and $\Lambda_c/E_{UV}$. The colored part of the figures represent
the allowed region, where all the constraints are satisfied. 
We also plotted two lines delimiting the  
weakly coupled regime of the conformal window ($2 \tilde N > (N_f^{(1)}+N_f^{(2)})$)
and the IR free window ($3 \tilde N < (N_f^{(1)}+N_f^{(2)})$).

From the figures we see that smaller values of the ratio $\epsilon_{UV}$ 
guarantees that the running can be longer in the CFT window.
The red region shaded in the figures,
near $N_f^{(1)} + N_f^{(2)}= 3 \tilde N$, 
is filled also if the running is extended over
a large regime of scales. At the lower edge of this region 
the  anomalous dimensions are close to zero,  the 
UV hierarchy imposed on the relevant deformations is preserved  during the flow,
and $\epsilon_{UV} \sim \epsilon_{IR}$.
As we approach the strongly coupled region of the conformal window
the anomalous dimensions
get larger.  In this case $\epsilon_{IR}$ approaches to one and we
represented this behavior by changing the color of the shaded region 
from red to orange and then to yellow. The white part of the figures represents the 
region in which $\epsilon_{IR}>1$.

In conclusion we have found regions in the parameter space where the
theory possesses metastable non supersymmetric vacua. The RG flow analysis
gives non trivial constraints on the relevant deformations and on the
duration of the approximate conformal regime.

\section{General strategy}
\label{dafare}

We discuss here the generalization of the mechanism of
supersymmetry breaking in SCFTs deformed by relevant operators. 
As in  SSQCD, the lifetime of the metastable vacuum
can be long  in the conformal window of other models,
with opportune choices of the parameters.
Consider a $SU(N_c)$ gauge theory with $N_f^{(1)}$ flavors
of quarks 
in the magnetic IR free window
and with  a
metastable supersymmetry breaking vacuum in the 
dual phase.
In the magnetic phase a new set of $N_f^{(2)}$ massive quarks 
must be added
to reach
the conformal window.
If there is some gauge invariant operator $\mathcal{O}$
that hits the unitary bounds, $R(\mathcal {O}) <2/3$, it is necessary to
add other singlets and also marginal couplings in the superpotential
between
the quarks and these new singlets.
The mass term for the new quarks
is a relevant perturbation which grows in the infrared,
and it has to be very small with respect to the other scales of the theory,
down to the CFT exit scale.
This mass term modifies the non perturbative superpotential
and the supersymmetric vacuum, which sets the CFT exit scale.
One must inspect a regime of couplings
such that  the supersymmetric vacuum is far away in the field space. 
This regime corresponds to a bound on the parameters of the theory, which 
have to be consistent with the RG running of the physical
coupling constants.  In the canonical basis  the running of the physical couplings
can be absorbed in the superpotential 
by the wave function renormalization of the fields.
If there is a relevant operator $\Delta W =\eta \mathcal{O}$, with classical dimension 
dim$(\mathcal{O})=d$, the physical coupling $\eta$ runs from the 
UV scale $E_{UV}$ to the IR scale $E_{IR}$ as
\be \label{importante}
\eta (E_{IR}) = \eta(E_{UV}) Z_{\mathcal{O}} (E_{IR},E_{UV})^{-\frac{1}{2}}=
\eta(E_{UV}) \left(\frac{E_{IR}}{E_{UV}}\right)^{\gamma/2}
\ee
We require that the running in this approximate conformal regime stops 
at the energy scale $\Lambda_c$ set by 
the masses at the supersymmetric vacuum.
The bounds on the parameters that ensure the stability of the metastable
vacuum have to hold at this IR CFT exit scale.  The equation 
(\ref{importante}) translates these bounds in some requirements 
on the UV deformations.
The metastable vacua have long lifetime if there is some
regime of UV couplings in which 
the stability requirements  are satisfied  in the weakly coupled 
conformal window.

Here we have shown that in SSQCD there are some regions in the conformal window in
which a large hierarchy among the couplings allows the existence of long
living metastable vacua.
We expect other models with this behavior.

\section{Discussion}
\label{disco}

In this paper we discussed the realization of the ISS mechanism in the conformal window 
of SQCD-like theory.
In \cite{Intriligator:2006dd} the metastable vacua disappeared
if $3/2N_c <N_f< 3N_c$ because the non perturbative dynamics was not negligible
in the small field region, and this destabilized the non 
supersymmetric vacua.

We have reformulated this problem in terms of the 
RG flow from the $UV$ cut-off of the theory down to the CFT 
exit scale.  
In the ISS model the CFT exit scale 
and the supersymmetry breaking scale are proportional 
because of the equation of motion of the meson.
Their ratio depends only on the gauge coupling constant 
at the fixed point.
The bounce action is proportional to this
ratio and cannot be parametrically long.

This behavior suggests a mechanism to evade the problem
and to build models with long living metastable vacua in 
the conformal window of SQCD-like theories.
A richer structure of relevant deformations than in the ISS
model is necessary.
Metastable vacua with a long lifetime can exist
if the bounce action
at the CFT exit scale 
depends on the relevant deformations and
it is not RG invariant.
We have studied this mechanism in an explicit model, the SSQCD,
and we have found that in this case, by adding a new mass term
for some of the quarks, the bounce action has a parametrical dependence 
on the relevant couplings. 
The RG flow of these couplings for different regimes of scales
sets the desired regions of $UV$ parameter that gives a large bounce action
in the $IR$.
We restricted the analysis to a region of ranks in which 
the model is interacting but weakly coupled, and the 
perturbative analysis at the non supersymmetric state is applicable.
It is possible to extend this example to other SCFT theories
as we explained in Section \ref{dafare}. 

It would be interesting to find some dynamical mechanism to explain 
the hierarchy among the different relevant
perturbations, that are necessary for the stability of the metastable vacua. For example 
in the appendix we see that in quiver gauge theories
the mass of the new quarks can be generated with a stringy instanton  as in \cite{ Argurio:2007qk,Aharony:2007db}.
The supersymmetry breaking metastable vacua that we have found in the conformal
window might be used in conformally sequestered scenarios, 
along the lines of \cite{Schmaltz:2006qs}.
Another application is the study of Yukawa interactions along the lines of
\cite{Nelson:2001mq,Nelson:2000sn}. Superconformal field theories
naturally explain the suppression of the Yukawa couplings if some
of the gauge singlet fields are identified with the $T_i = 10_i$ and $F_i = \bar 5_i$ 
 generations of the $SU(5)$ GUT group. Here we have shown that supersymmetry breaking in superconformal sectors is viable. 
It is in principle  possible to build a supersymmetry breaking SCFT where some 
of the generation of the MSSM 
are gauge singlets, marginally interacting with the fundamentals of the SCFT group. 
In this case the Yukawa arising from these generations can be suppressed as in
\cite{Nelson:2001mq,Nelson:2000sn}. 
Since supersymmetry is broken one can imagine a mechanism of flavor blind mediation,
like gauge mediation, to generate the soft masses for the rest of the multiplets of the MSSM. 
Closely related ideas has recently appeared in \cite{Poland:2009yb} and \cite{Aharony:2010ch} .

\section*{Acknowledgments}

We are grateful to Kenneth Intriligator for valuable and stimulating
discussions.  We also thank Riccardo Argurio, Jeff Fortin, Sebastian
Franco, Riccardo Rattazzi and Angel Uranga for comments.
M.S.  also thanks Silvia Penati for useful discussions on the manuscript.

A.~ A.~ is supported by UCSD grant DOE-FG03-97ER40546.
L.~G.~ and M.~S.~ are supported in part by INFN,
in part  by
MIUR under contract 2007-5ATT78-002.
A. ~M. ~ is a 
Postdoctoral researcher of FWO-Vlaanderen.
A. ~M. ~ is also
supported in part 
by the Belgian Federal Science Policy Office 
through the Interuniversity 
Attraction Pole IAP VI/11 and by 
FWO-Vlaanderen through project G.0428.06.

\appendix
\section{The renormalization of the bounce action}
\label{appbounce}
In the paper we analyzed the bounce action at the 
CFT exit scale. We distinguished the infrared 
bounce action $S_{B,IR}$
from $S_{B,UV}$, the action evaluated at the UV scale.
Indeed in a supersymmetric field theory 
in the holomorphic basis
the bounce action
is obtained from the Lagrangian
\be
\mathcal{L} = Z_{\phi} \dot \phi^2 + Z_{\phi}^{-1}V(\phi)
\ee 
and we have
\be\label{Zbo}
S_{B,IR} = S_{B,UV} Z_{\phi}^{3}
\ee
Hence the bounce action undergoes non trivial renormalization.
Here we show that our analysis, performed in the canonical basis,
is consistent with (\ref{Zbo}), both for the ISS model and for the model
in Section \ref{mainsec}.

The ISS bounce action in the UV is 
\be
S_{B,UV} = \left(
\frac{\mu_{UV}}{\tilde \Lambda_{UV}}
\right)^{\frac{4 \tilde b}{N_f-\tilde N}}
\ee
In the IR this action is renormalized because of the 
wave function renormalization of the fields.
In the paper we computed the action in the canonical basis and 
renormalization effects have been absorbed into the couplings. 
From (\ref{Zbo}) the IR renormalized action $S_{B,IR}$ is
\be \label{SBIR}
S_{B,IR} = S_{B,UV} Z_{N}^{3} 
\ee
where the wave function renormalization is 
\be
Z_M = \left( \frac{E_{IR}}{E_{UV}} \right)^{-\gamma_N}
\ee
We now compute $S_{B,IR}$ and show that indeed it is (\ref{SBIR}).
The coupling $\mu_{IR}$ and the  
scale $\tilde \Lambda_{IR}$ are given as functions of their UV values
\be
\mu_{IR} = \mu_{UV} Z_N^{-1/4},
\qquad \qquad
\tilde \Lambda_{IR} = \tilde \Lambda_{UV} \frac{E_{IR}}{E_{UV}}=
\tilde \Lambda_{UV} Z_N^{-1/\gamma_N}
\ee
By substiting on the l.h.s. of 
(\ref{SBIR}) we have
\be \label{mianonna}
S_{B,IR} = \left(\frac{\mu_{IR}}{\tilde \Lambda_{IR}} \right)^{\frac{4 \tilde b}{N_f-\tilde N}}
=\left( \frac{\mu_{UV} Z_N^{-1/4}}{\tilde \Lambda_{UV} Z_N^{-1/\gamma_N}}
\right)^{\frac{4 \tilde b}{N_f-\tilde N}}=
S_{B,UV} Z_N^{3}
\ee
where the last equality is obtained by substituting $\tilde b=2N_f-3N_c$
and $\gamma_N= 2 \tilde b/N_f$.
Nevertheless the bounce action in SQCD at the CFT exit scale results RG invariant. This is
because the mass scales of the theory are related by  the equation
of motion of $N$. The relation 
between these scales is proportional to the gauge coupling which is constant
during the running in the conformal window.
For this reason the lifetime of the metastable vacuum cannot be parametrically large in SQCD.

In the model discussed in section \ref{mainsec} instead the bounce action
depends non trivially on the relevant deformations
\begin{equation}
S_{B,IR} = \left(\frac{\tilde \Lambda_{IR}}{\rho_{IR}}\right)^{\frac{4 N_f^{(2)}}{N_f^{(1)}-\tilde N }       } 
\left(\frac{\mu_{IR}}{\tilde \Lambda_{IR}}\right)^{\frac{ 12 \tilde N - 4N_f^{(1)}}{N_f^{(1)}-\tilde N}  } 
\end{equation}
The UV bounce action has the same expression but in term of the UV couplings and scale.
The IR coupling and scale are related to the UV values as
\be
\mu_{IR}=\mu_{UV} Z_N^{-1/4} \qquad \rho_{IR}=\rho_{UV} Z_{N}^{-\frac{\gamma_p}{\gamma_N}}
\qquad \tilde \Lambda_{IR}=\tilde \Lambda_{UV} Z_N^{-\frac{1}{\gamma_N}}
\ee
The infrared bounce action is then
\be
S_{B,IR}=S_{B,UV} Z_N^A
\ee
where
\be
A=\frac{4 N_f^{(2)}(\gamma_p-1)  }{(N_f^{(1)} -\tilde N)\gamma_N  } 
+\frac{(3 \tilde N-  N_f^{(1)})(4- \gamma_N)}{(N_f^{(1)} - \tilde N) \gamma_N}=3
\ee
The last equality can be obtained by substituting the relations 
$\gamma_{\phi_i}= 3 R[\phi_i]-2$, with
 $R[N]=2y$ and $R[p]=(n-x+y)/n$. 
Hence we verified the general result (\ref{Zbo}) concerning the renormalization of the bounce action.

\section{The SSQCD}
\label{appanto}

In this appendix we review the SSQCD 
defined in \cite{Barnes:2004jj} and its behavior 
under Seiberg duality.
The model is a $SU(N_c)$ gauge theory with quarks charged under the
 $SU(N_f^{(1)})\times SU( N_f^{(2)})$
flavor symmetry  and a singlet in the bifundamental of $SU(N_f^{(2)})$.
The matter content is given in Table \ref{tabella2}.
\begin{table}[htbp]
\begin{center}
\begin{tabular}{c||c|c|c}
&$N_f^{(1)}$&$N_f^{(2)}$&$N_c$ \\
\hline 
$\tilde Q +  Q$&$\bar{N}_f^{(1)} \oplus N_f^{(1)}$&1&$N_c \oplus \bar{ N}_c$\\
$\tilde P + P$&1&$\bar{N_f}^{(2)}\oplus N_f^{(2)}$&$ N_c \oplus \bar{ N}_c$\\
$S$&1&$\bar{N_f}^{(2)} \otimes N_f^{(2)}$&1\\
\end{tabular}
\caption{Matter content of the SSQCD}
\label{tabella2}
\end{center}
\end{table}
The superpotential is 
\be
W = S P \tilde P
\ee
In the conformal window, $3/2 N_c < N_f^{(1)} + N_f^{2)} < 3 N_c$
there is a Seiberg dual description, with 
$SU(N_f^{(1)}+ N_f^{(2)}- N_c)$  magnetic gauge group with 
matter content given in Table \ref{tabella1}, where the 
mass term for the field $S$ and the meson
$M = P \tilde P$ is integrated out.
The dual superpotential is
\be\label{WVERG}
W = K p \tilde q + L \tilde p q + N q \tilde q 
\ee
In the conformal window these theories are dual if
there are no accidental symmetry, not manifest 
in the UV Lagrangian, that emerges in the IR. 

If some accidental symmetry arise, some  gauge invariant operator,
$\mathcal{O}$, 
in the chiral ring, violates the unitary bound and we have
$R(\mathcal{O})<2/3$ from the a-maximization. 

The marginal term in the superpotential associated to this
operator becomes irrelevant and can be neglected in the IR.

In SSQCD the first operator that hits the unitary bound is $N = Q \tilde Q$. 
By using the $y_{max}$ that we calculated in (\ref{cicciogay})  
we see that the unitary bound is hit at 
\be
x =  \frac{1}{3} \left(2-2 n+\sqrt{1-14 n+13 n^2}\right)
\ee
where $x=\frac{\tilde N}{N_f^{(1)}}$ and $n=\frac{N_f^{(2)}}{N_f^{(1)}}$.
For higher values of $x$ the dual superpotential becomes 
\be
W = K p \tilde q + L \tilde p q 
\ee
In the paper we have studied a region were this meson does not
hit the unitary bounds, and we can trust the duality without adding new operators.
\begin{figure}[htbp]
\begin{center}
\includegraphics[width=10cm]{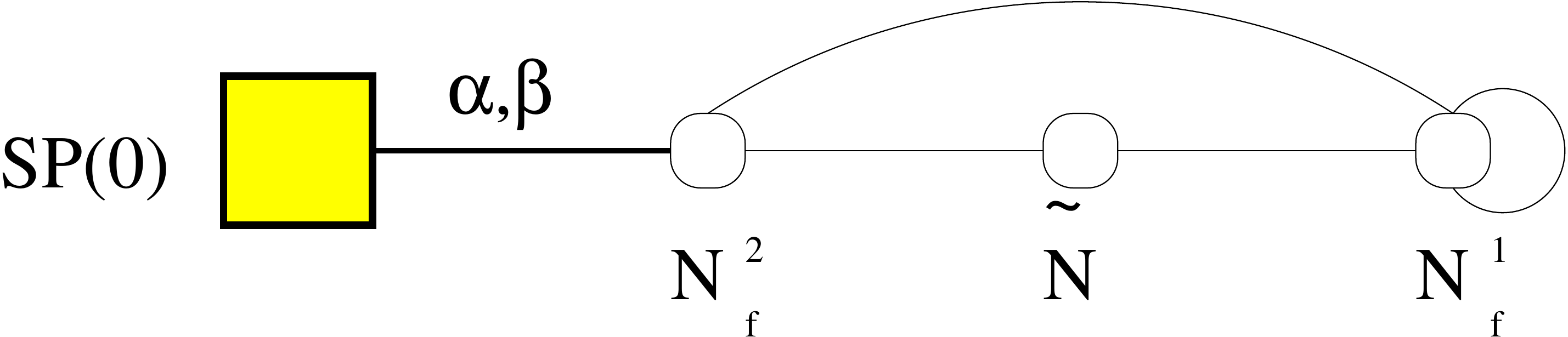}
\end{center}
\caption{The Stringy instanton contribution}
\label{mivergogno}
\end{figure}
\subsection*{Relevant deformations}

Some deformations must be added to (\ref{WVERG}) to recover (\ref{spotimpo}).
The linear term for $M$ can be generated in the electric gauge theory
by adding a mass term for the quarks $Q$ and $\tilde Q$,
while the mass term for the field $p$ and $\tilde p$ can be generated
by adding a linear deformation $k^2 S$.
When we integrate out the mass term $m M S$ in the magnetic theory
the fields  $p$ and $\tilde p$ acquire a mass term proportional to 
$\rho= k^2/m$.

However a large hierarchy is required between the scale $\mu$ and the mass $\rho$ for the existence of the metastable vacua. We can impose this hierarchy
at hand or find a dynamical mechanism.
For example, when $N_f^{(2)}=1$ we can think to embed the magnetic
theory in a quiver 
and couple the fields $p$ and $\tilde p$ with an $SP(0)$ node as in Figure 
\ref{mivergogno}

In the instantonic action an interaction $ S \sim \alpha p \tilde p \beta $
between the instanton moduli and the fields is present.
By integrating over the instantonic zero modes we are left with the desired  suppressed
mass term
\be
\Delta W =\int \text{d}\,\alpha \,\text{d}\,\beta\,\, e^{S_{inst}}= \Lambda e^{-A} p \tilde p
\ee
for the $p$ and $\tilde p$ quarks, where A represents the area 
of curve associated to the $SP(0)$ node
and $\Lambda$ is associated to a string scale.

  \end{document}